\documentstyle[11pt,paspconf,psfig]{article}

\renewcommand{\thepage}{\arabic{page}}
\begin{document}

\title{Profile variations in AGN spectra}

\author{W.\ Kollatschny and K.\ Bischoff}
\affil{Universit\"ats-Sternwarte, Geismarlandstr.\ 11,
D-37083 G\"ottingen, Germany}

\begin{abstract}
We present results of optical long-term variability campaigns (10 -- 20 years)
of the two Seyfert galaxies NGC~7603 and Mrk~110.   The variations of the
continuum, of  the individual broad line intensities and of their line profiles
are investigated in detail and compared to line profile variations in NGC~5548
and NGC~4593. Individual emission line profiles vary differently  from line
to line and from outburst to outburst indicating a complex and structured
broad emission line region.
\end{abstract}



\section{Introduction}

Many Seyfert~1 galaxies are known to be variable in the
 continuum and in the broad emission
line fluxes on time scales of weeks to years.
They might be variable on even shorter time scales. The 
continuum and line intensity variations can give us information on
the radius of the broad line region (BLR) and their
internal kinematics. 


We will present some results on line and continuum variations in
the Seyfert galaxies NGC~7603 and Mrk~110. These 
results will be compared to earlier variability campaigns of NGC~5548
(AGN Watch: Peterson et al., 1991; Kollatschny \& Dietrich 1996) 
 and NGC~4593 (LAG campaign: Robinson 1994;
Kollatschny \& Dietrich 1997).

We are interested to study the variations of as many broad emission lines per
galaxy as possible to look for individual differences in their amplitudes and
lags with respect to continuum variations and to look  for internal profile
changes.

\section{Observations} 

We obtained optical spectra of NGC~7603
 over a long period from
1979 until 1996 with typical sampling intervals
 of one year. NGC~7603 is a bright
Seyfert~1 galaxy ($\rm M_V=-21.5$) with broad line profiles
 (FWHM(H$\beta$)=6500~km~s$^{-1}$). 
Most of our spectra covering the whole optical spectral range (see Figure~1)
were  taken at Calar Alto Observatory and at McDonald Observatory.
\begin{figure}
\vspace{0.cm}
\psfig{figure=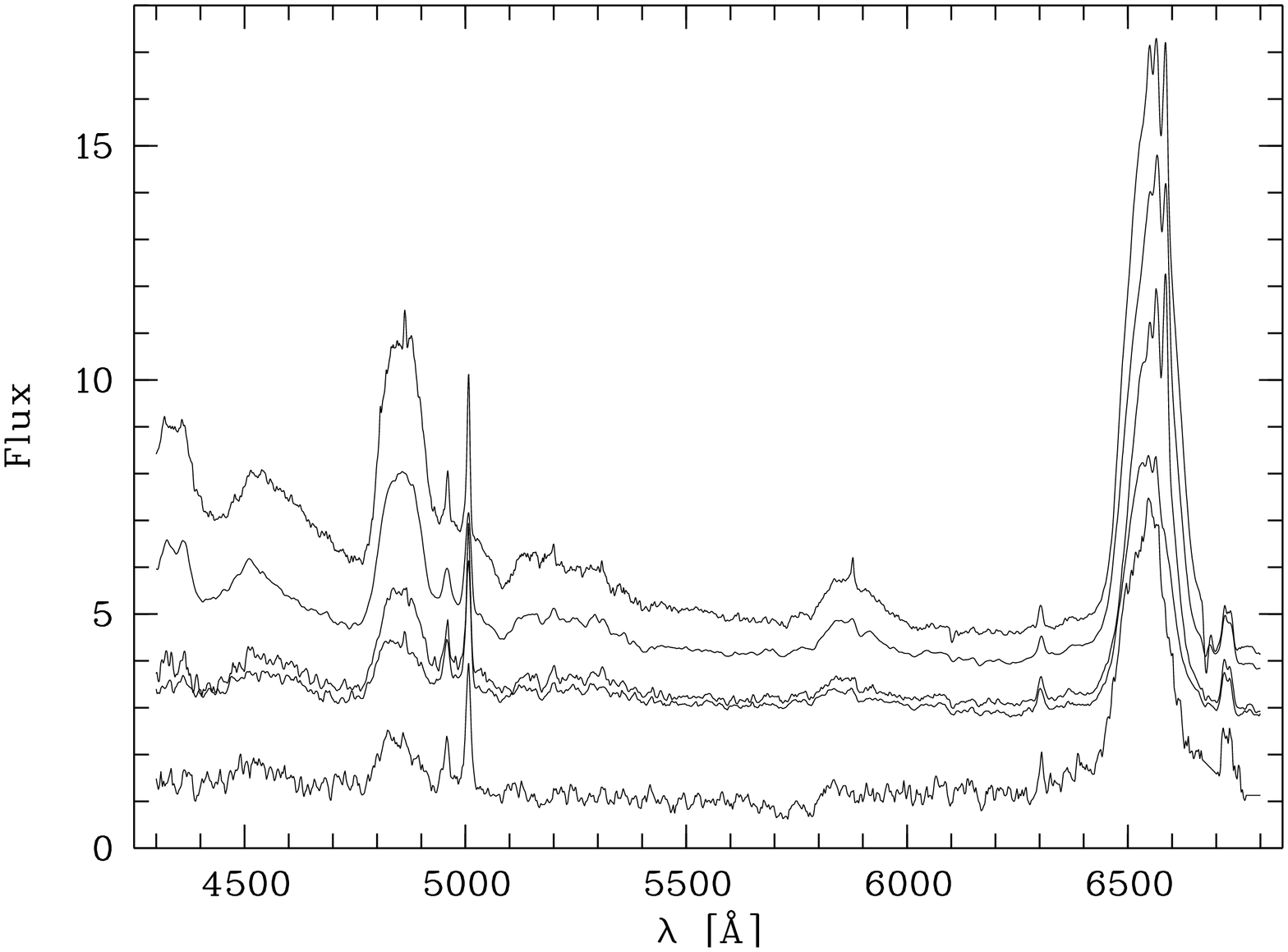,angle=270,height=9cm}
\caption{Normalized spectra of NGC~7603 taken at different epochs
during the years 1979, 1988, 1990, 1992, 1993 (from bottom to top)}
\vspace{0.cm}
\psfig{figure=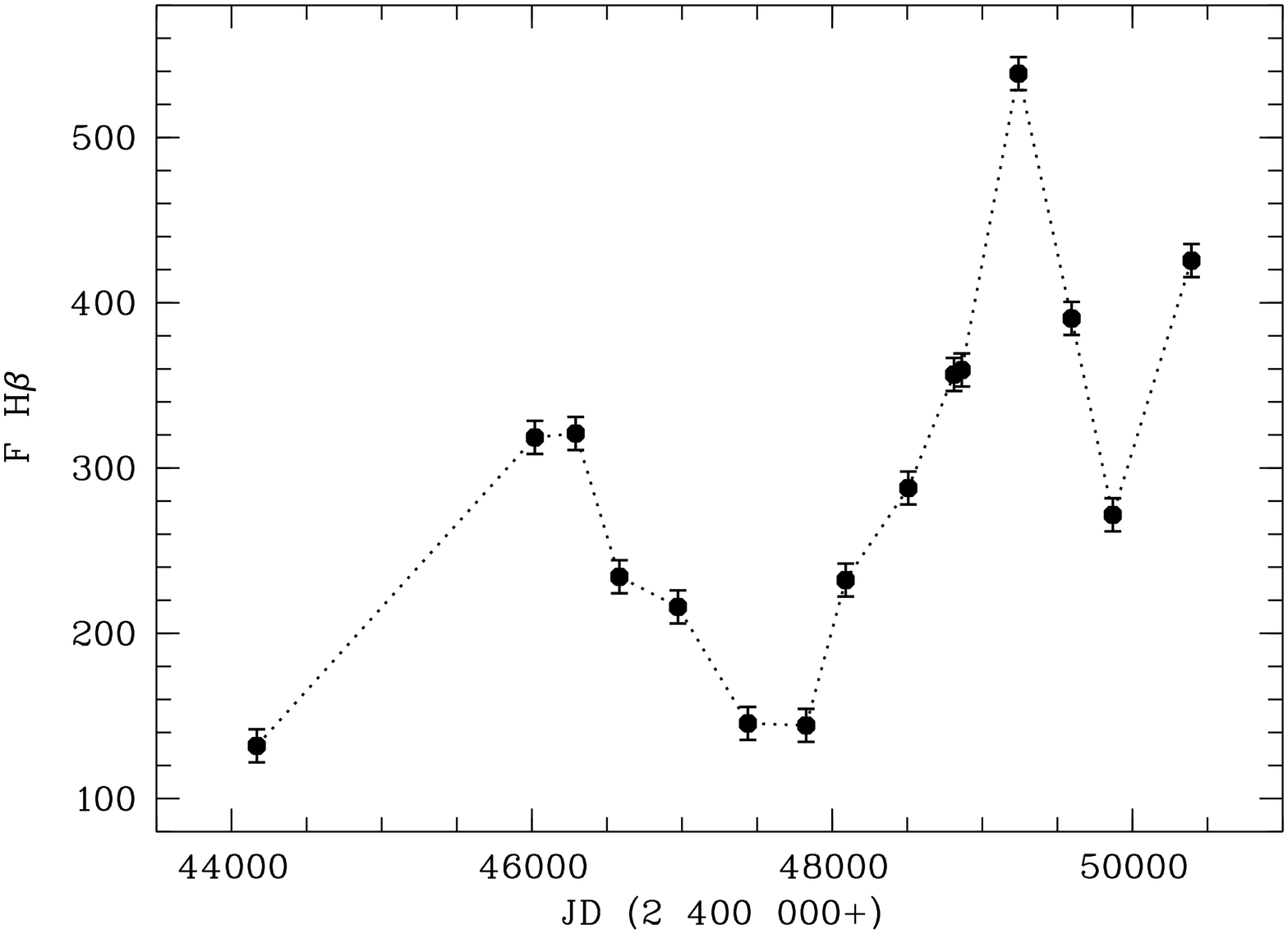,angle=270,height=9cm}
\caption{H$\beta$ light curve of NGC~7603 from 1979 until 1996. The numbers
 give the last digits of the Julian Date.} 
\end{figure}

Spectra of Mrk~110 ($\rm M_V$=-20.6) have been taken
 at the same observatories from
1987 until 1995 at 24 epochs.  
The emission line profiles of Mrk~110 are quite narrow
 (FWHM(H$\beta$)=1800~km~s$^{-1}$) similar
to the so called Narrow Line Seyfert Galaxies.
\begin{figure}
\vspace{0.cm}
\psfig{figure=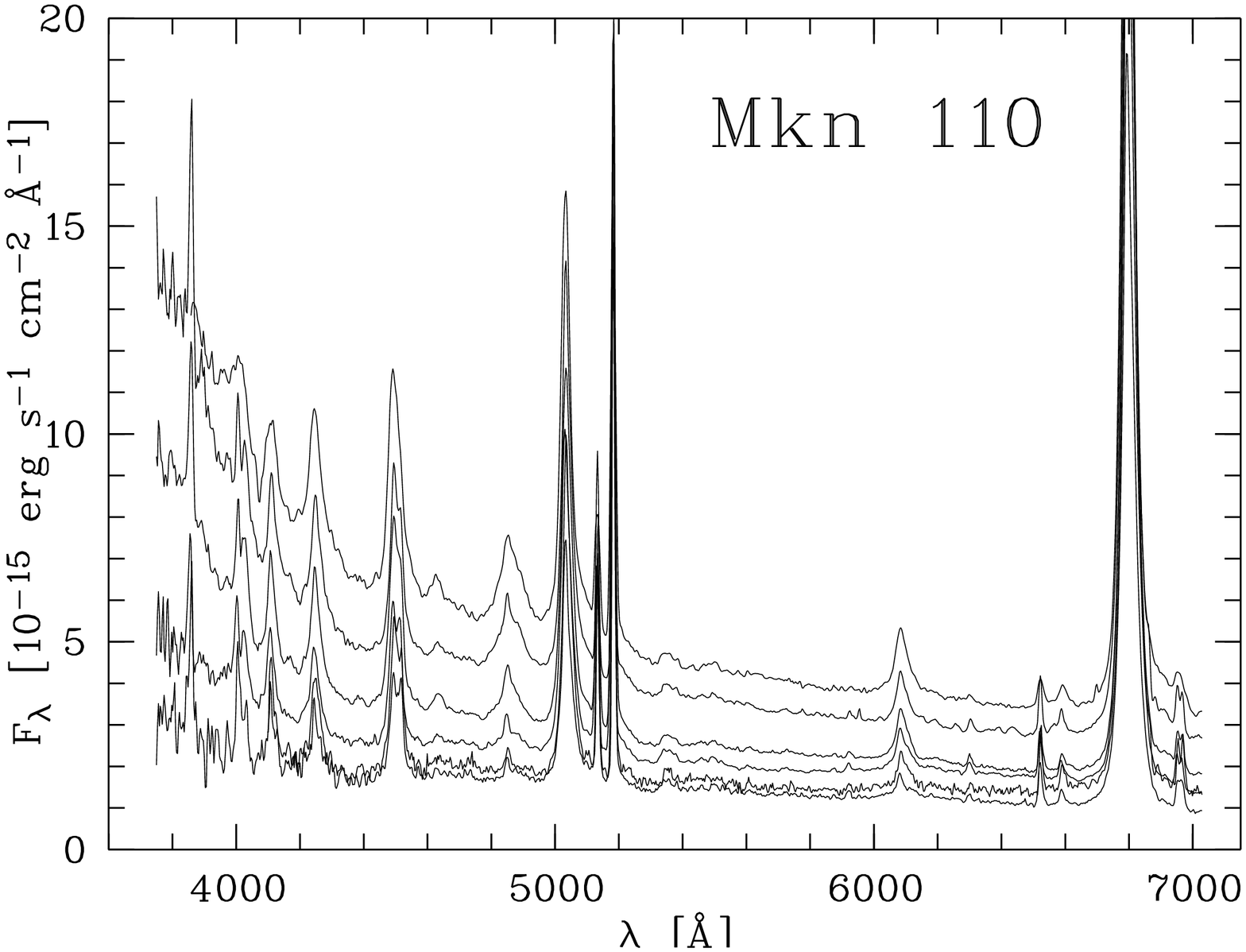,angle=270,height=9cm}
\caption{Normalized spectra of Mrk~110 taken at different epochs
during 1988, 1989, 1988, 1989, 1987, 1989, 1992
(from bottom to top).} 
\vspace{0.cm}
\psfig{figure=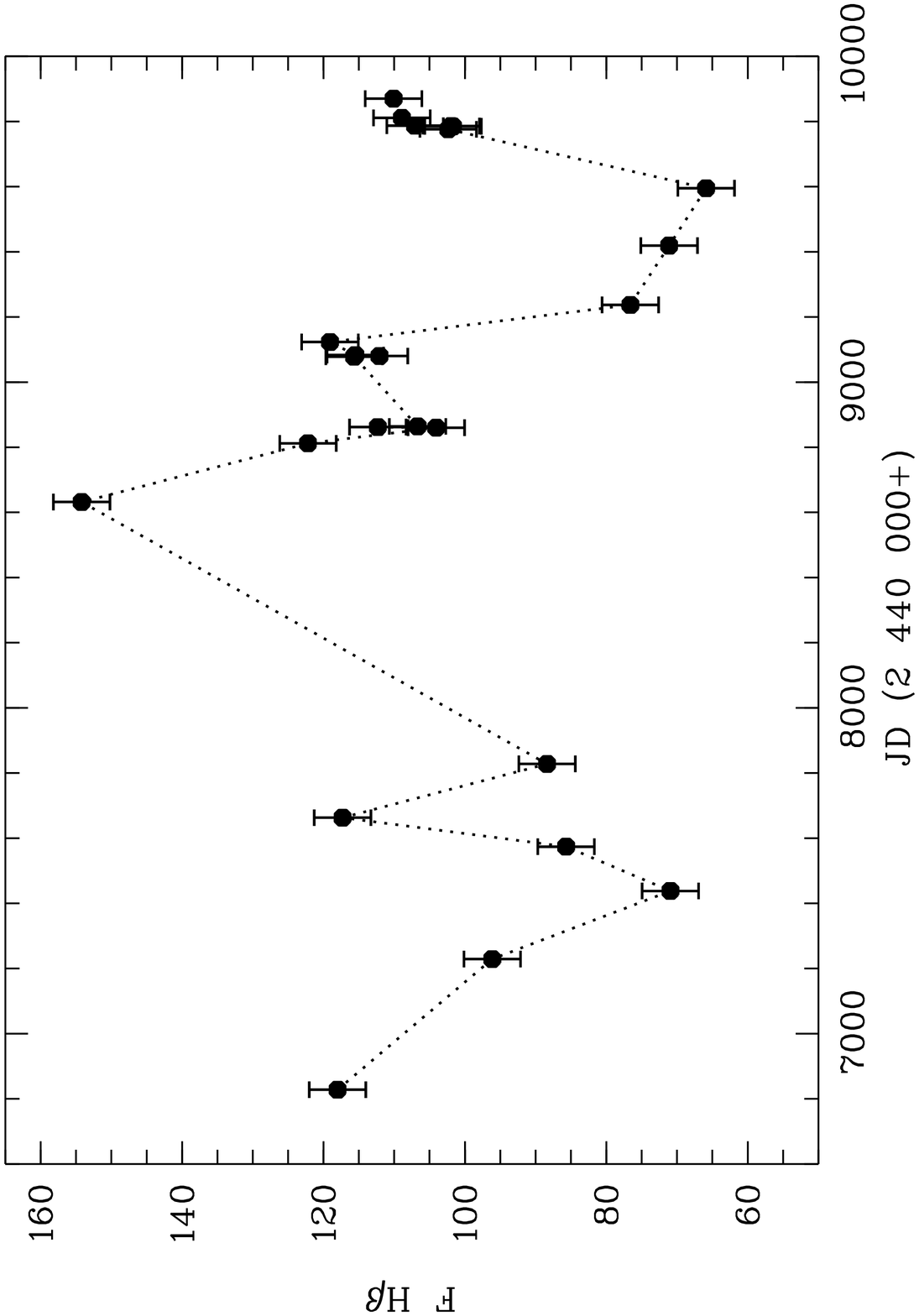,angle=270,height=9cm}
\caption{H$\beta$ light curve of Mrk~110 from 1987 until 1995.
The numbers give the Julian Date.}
\end{figure}

NGC~5548 has been monitored in the optical by many observers
with different telescopes 
 from December 1988 until October 1989 during the first year of the
AGN Watch (Peterson et al. 1991, Kollatschny \& Dietrich 1996). High quality
H$\alpha $, H$\beta $, H$\gamma $, He\,{\sc i}, and He\,{\sc ii}
 spectra have been obtained at 30 to 70 epochs.

H$\alpha $ and H$\beta $
spectra of NGC~4593 were taken at La Palma with the WH and IN
telescopes from January
until June 1990 at 23 epochs as part of the LAG campaign.

\section{Results and Discussion}

Some of our
optical spectra of NGC~7603, normalized to the narrow forbidden lines,
are shown in Figure~1. They represent
 five different epochs at low, intermediate and high stages.
Our observed H$\beta$ light curve is plotted for the period
 1979 until 1996 in Figure~2.
Strong variations in the emission line intensities and profiles
are clearly to be seen.
The broad Fe\,{\sc ii} multiplets at $\lambda\lambda$4400--4700 \AA\  and
$\lambda\lambda$5080--5400 \AA\  show variations of the same order. Our
light curve of NGC~7603 was not sampled with high frequency. Therefore,
the cross correlation analysis (CCF) of the emission line intensities 
with the continuum light curve gives only a coarse estimate of the
 radius of the BLR of the order of 50 light days.

Normalized spectra of Mrk~110 are plotted in Figure~3.
The H$\beta$ light curve over the period from 1987 until 1995
 is given in Figure~4.
The highly excited He\,{\sc ii}~$\lambda$4686 line
in Mrk~110 shows the largest amplitude
in the light curve, the broadest line profile and the strongest profile
changes. Indication of strong variability in the He\,{\sc ii} line has already
been mentioned by Peterson (1988).
Our CCF analysis reveals that the broad He\,{\sc ii} line originates close to the
central ionizing source (in only 6 light-days distance) in comparison
to the Balmer lines (40--70 light-days).
\begin{figure}
\vspace{0.cm}
\psfig{figure=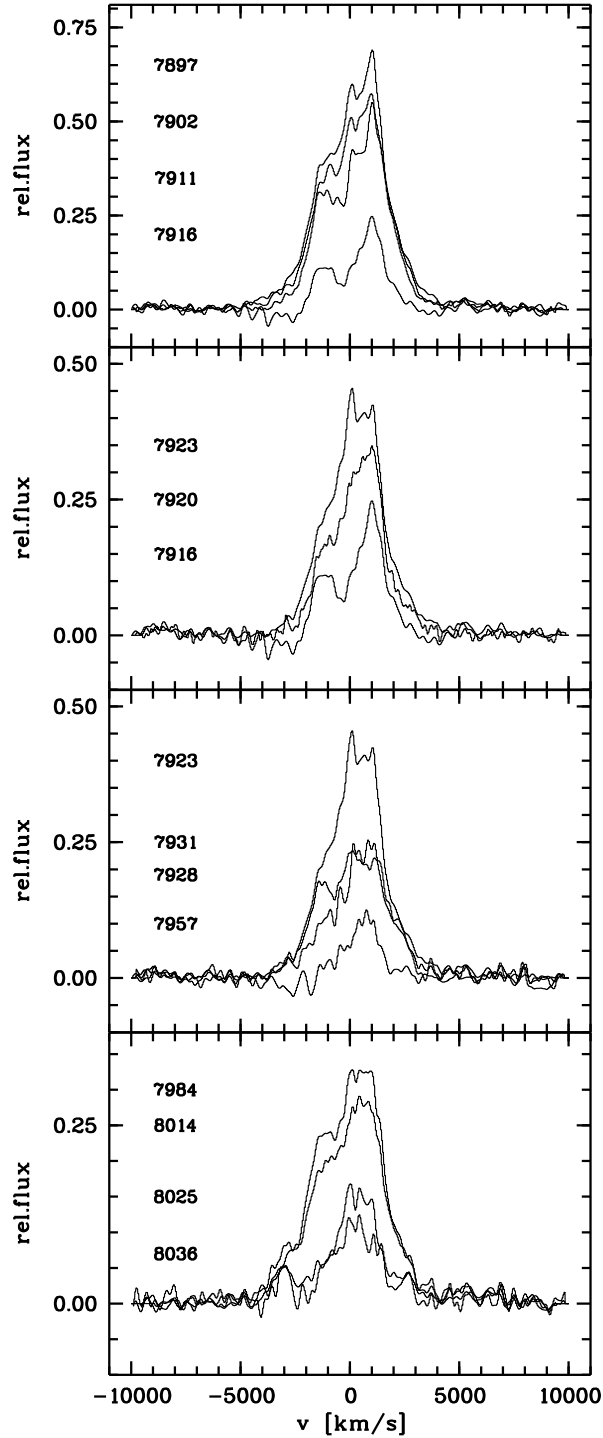,angle=0,width=8cm,clip=}
\caption{NGC~4593: Difference profiles of H$\alpha$ with respect to the minimum
spectrum transformed into velocity space. The numbers give the last digits
of the Julian Date (2440000+).}
\end{figure}
\begin{figure}
\vspace{0.cm}
\psfig{figure=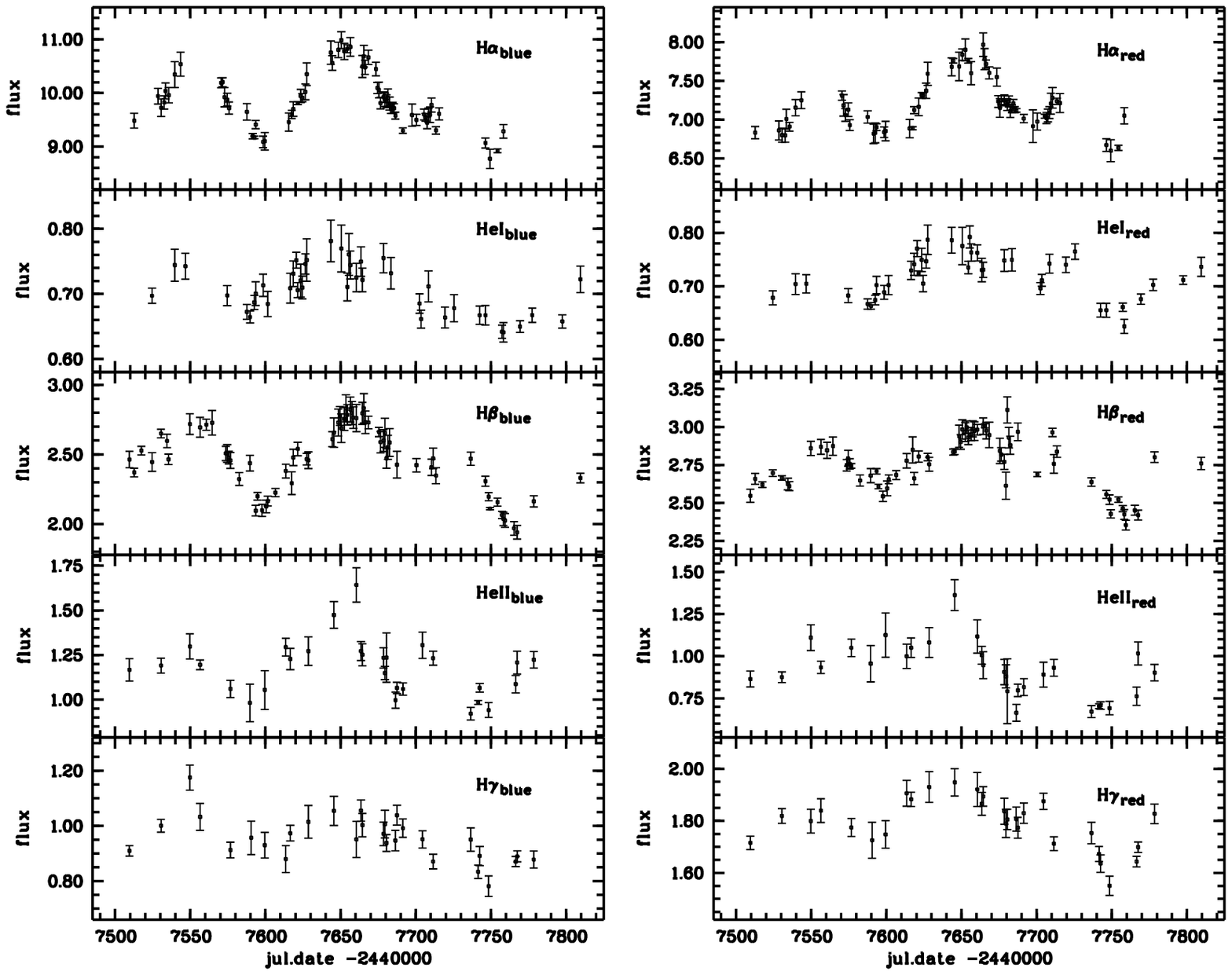,angle=0,height=9.cm,clip=}
\caption{Individual light curves of the blue and red wings of H$\alpha$,
He\,{\sc i}~$\lambda$5876, H$\beta$, He\,{\sc ii}~$\lambda$4686 and H$\gamma$ of
NGC~5548 during the year 1989.}
\end{figure}

Detailed analyses of line profile variations in NGC~5548 and NGC~4593
have been published earlier by us (Kollatschny \& Dietrich 1996, 1997).
Individual H$\alpha$ difference profiles with respect to our
minimum state of NGC~4593 are plotted in Figure~5.
There are variations on time scales of days only. The red
and blue wings vary differently. Individual segments in the profiles
appear and disappear on short time scales.

 Light curves of the blue and red line wings of different
emission lines in NGC~5548 are given in Figure~6. They differ with respect 
to each other and from line to line.

The observational results of the BLR are very complex.
 But one can make some
general statements.
The higher ionized line have broader line profiles, show larger variability
amplitudes and origin closer to the central ionizing source in comparison
to the lower ionized lines. However, the
 Fe\,{\sc ii} lines in NGC~7603 show strong variations, too.

There are line profile variations on time scales of days.
The emission line cores and wings vary differently. Line
asymmetries are different from line to line and from outburst to outburst.
There is a trend that the line profiles become more symmetric with increasing 
line flux.
 The cross correlation functions of the
 light curves of individual line segments
with the continuum light curves showed
 that in some cases the outer line wings respond
faster. But the response is different from line to line
during the same outburst.
Strong radial motions can be excluded to be dominant in the BLR. The BLR
seems to be clumpy and to consist of a limited number of clouds or cloud
complexes only. These clouds or the ionizing source might have a
bidimensional structure.

\acknowledgments

This work has been supported by Deut\-sche Agentur f\"ur
 Raumfahrtangelegenheiten (DARA) grant 50\,OR\,9408\,9 and
 DFG grant Ko857/13.

\newpage
\begin{question}{Bill Welsh}
Have you compared the r.m.s. and mean line profiles, especially with respect
to the maximum width of the line~(FWZI)?
\end{question}
\begin{answer}{Wolfram Kollatschny}
In many cases (e.g. NGC~7603) the widths of the mean and r.m.s. profiles are
identical. However, in other cases (He lines in NGC~5548) the widths
are completely different. 
\end{answer}

\begin{question}{Bill Welsh}
Over such long time scales (20 years), do you still expect the NLR lines (e.g. 
[O\,{\sc iii}]) to remain constant and therefore usable for internal
 self--calibration of the fluxes?
\end{question}
\begin{answer}{Wolfram Kollatschny}
We verified on narrow band [O\,{\sc iii}] images that the NLR is a point source.
But we can not exclude that the NLR might be variable on the order of
a few percent over a time scale of 20 years.
\end{answer}

\begin{question}{Brian Espey}
For NGC~7603 you showed that there are substantial Fe\,{\sc ii} intensity variations.
What proportion of the H$\beta$ variability is due to variations in the
underlying Fe\,{\sc ii} multiplet strengths?
\end{question}
\begin{answer}{Wolfram Kollatschny}
>From Figure~1 one can estimate that the blending
due to Fe\,{\sc ii} multiplets corresponds to 10--30 percent of the H$\beta$
intensity. 
\end{answer}

\end{document}